\documentclass[letterpaper,twocolumn,prl,aps,superscriptaddress,amsmath,amssymb,floatfix]{revtex4-1}
\usepackage{color}
\usepackage{babel}
\usepackage{amsmath}
\usepackage{graphicx}
\usepackage[unicode=true,pdfusetitle,
 bookmarks=true,bookmarksnumbered=false,bookmarksopen=false,
 breaklinks=false,pdfborder={0 0 1},backref=false,colorlinks=true]
 {hyperref}
\hypersetup{
 pdfborderstyle=,colorlinks,linkcolor=red,citecolor=blue}

\usepackage{soul}



\begin{document}
\affiliation{CAS Key Laboratory of Quantum Information, University of Science and Technology
of China, Hefei, Anhui 230026, P. R. China}
\affiliation{Hefei National Laboratory,
University of Science and Technology of China, Hefei 230088, China}
\affiliation{Department of Chemical Physics, University of Science and Technology of China, Hefei 230026, China}
\affiliation{State Key Laboratory of Quantum Optics and Quantum Optics Devices,
and Institute of Opto-Electronics, Shanxi University, Taiyuan 030006,
China}
\affiliation{Institute of Advanced Science Facilities, Shenzhen, 518107, China}

\global\long\def\figurename{FIG.}%
\title{
Fano-like resonance due to interference with distant transitions}
\author{Y.-N. Lv} 
\thanks{These authors contributed equally to this work.}
\affiliation{CAS Key Laboratory of Quantum Information, University of Science and Technology
of China, Hefei, Anhui 230026, P. R. China}
\affiliation{Hefei National Laboratory,
University of Science and Technology of China, Hefei 230088, China}
\author{A.-W. Liu} 
\thanks{These authors contributed equally to this work.}
\affiliation{Hefei National Laboratory,
University of Science and Technology of China, Hefei 230088, China}
\affiliation{Department of Chemical Physics, University of Science and Technology of China, Hefei 230026, China}
\author{Y. Tan} 
\thanks{These authors contributed equally to this work.}
\affiliation{Hefei National Laboratory,
University of Science and Technology of China, Hefei 230088, China}
\affiliation{Department of Chemical Physics, University of Science and Technology of China, Hefei 230026, China}
\author{C.-L. Hu} 
\affiliation{Department of Chemical Physics, University of Science and Technology of China, Hefei 230026, China}
\author{T.-P. Hua} 
\affiliation{Department of Chemical Physics, University of Science and Technology of China, Hefei 230026, China}
\author{X.-B. Zou}
\affiliation{CAS Key Laboratory of Quantum Information, University of Science and Technology
of China, Hefei, Anhui 230026, P. R. China}
\affiliation{Hefei National Laboratory,
University of Science and Technology of China, Hefei 230088, China}
\author{Y. R. Sun}
\email{robert@ustc.edu.cn}
\affiliation{Hefei National Laboratory,
University of Science and Technology of China, Hefei 230088, China}
\affiliation{Department of Chemical Physics, University of Science and Technology of China, Hefei 230026,
China}
\affiliation{Institute of Advanced Science Facilities, Shenzhen, 518107, China}
\author{C.-L. Zou} 
\email{clzou321@ustc.edu.cn}
\affiliation{CAS Key Laboratory of Quantum Information, University of Science and Technology
of China, Hefei, Anhui 230026, P. R. China}
\affiliation{Hefei National Laboratory,
University of Science and Technology of China, Hefei 230088, China}
\affiliation{State Key Laboratory of Quantum Optics and Quantum Optics Devices,
and Institute of Opto-Electronics, Shanxi University, Taiyuan 030006,
China}
\author{G.-C. Guo} 
\affiliation{CAS Key Laboratory of Quantum Information, University of Science and Technology
of China, Hefei, Anhui 230026, P. R. China}
\affiliation{Hefei National Laboratory,
University of Science and Technology of China, Hefei 230088, China}
\author{S.-M. Hu} 
\email{smhu@ustc.edu.cn}
\affiliation{Hefei National Laboratory,
University of Science and Technology of China, Hefei 230088, China}
\affiliation{Department of Chemical Physics, University of Science and Technology of China, Hefei 230026, China}

\date{\today}

\begin{abstract}
Narrow optical resonances of atoms or molecules have immense significance in various precision measurements, such as testing fundamental physics and the generation of primary frequency standards. 
In these studies, accurate transition centers derived from fitting the measured spectra are demanded, which critically rely on the knowledge of spectral line profiles.
Here, we propose a new mechanism of Fano-like resonance induced by distant discrete levels 
 and experimentally verify it with Doppler-free spectroscopy of vibration-rotational transitions of CO$_2$.
The observed spectrum has an asymmetric profile and its amplitude increases quadratically with the probe laser power.
Our results 
facilitate a broad range of topics based on narrow transitions. 
\end{abstract}
\maketitle

Precision spectroscopy of narrow transitions of atoms and molecules has been the subject of numerous studies in recent decades and has been widely applied in sensing, metrology, and frequency references for optical clocks~\cite{Leopardi_2021_Metrologia,Brewer_2019_PRL,Seiferle_2019_Nature, McGrew_2018_Nature}. Narrow optical resonances also provide excellent probes for determining fundamental physics constants, such as the Rydberg constant~\cite{Parthey_2011_PhysRevLett} and the proton-to-electron mass ratio~\cite{Patra_2020_Science_HD, TAO_2018_PRL,Kortunov_2021_NP}. They are also candidates for investigating quantum electrodynamics and relativistic quantum mechanics and for exploring new physics and the dark matter~\cite{Cozijin_2018_PRL,Baron2014}. To circumvent the drawback of the ultraweak absorption of light for these transitions, strong laser fields are usually employed to probe these transitions. Moreover, counter-propagating laser beams, producing a standing-wave optical field, are often applied to reduce or eliminate the Doppler shift of molecules due to the motion along the laser beam~\cite{Fast_2020_PRL, VanRooij_2011_Science}. As a result, the spectra of narrow resonances in strong standing-wave fields have been studied extensively in recent decades.

In conventional atomic and molecular spectroscopy, 
two-level or few-level approximations are generally employed, and abundant far-off resonance excess transitions associated with the target energy levels are omitted in practice. Consequently,
\textit{symmetric} spectral profiles (mostly the Lorentz function) are assumed in practical 
precision measurements~\cite{Seiferle_2019_Nature,Brewer_2019_PRL,TAO_2018_PRL,Cozijin_2018_PRL,Patra_2020_Science_HD,Safronova_2018_RevModPhys}. 
It was surprising that an \textit{asymmetric} profile was observed in saturated absorption spectroscopy of isolated rotation-vibration transitions of HD while spectra of other molecules taken under the same experimental conditions remain symmetric~\cite{TAO_2018_PRL, Cozijin_2018_PRL, Diouf2019OL-HD, Hua2019}. 
The physics of this asymmetric profile has not been interpreted, 
which limits the investigation 
toward a determination of the dimensionless proton-to-electron mass ratio~\cite{TAO_2018_PRL}.
A notable difference between the asymmetric line of HD and the nearby symmetric lines of other molecules is that the HD line is extremely weak 
(Einstein coefficient $A \sim 10^{-5}$~s$^{-1}$).
Therefore, the puzzle may pose
challenges for evaluating the spectral data and testing fundamental physics at high precision~\cite{Cozijin_2018_PRL,Kozlov_2018_RevModPhys}, where various narrow transitions of atoms and molecules are involved. 

Here, we propose and experimentally demonstrate a new mechanism of nonlinear Fano-like resonance (NFR) among a few discrete energy levels in atoms or molecules, where  distant off-resonance but very strong transitions interfere with a near-resonance weak transition. %
The effect is numerically investigated and experimentally verified with vibration-rotational transitions of the CO$_2$ molecules.
This NFR mechanism can significantly change the shape of the spectrum and cause considerable deviations in precision measurements.


\begin{figure}
\centering{}\includegraphics[width=\columnwidth]{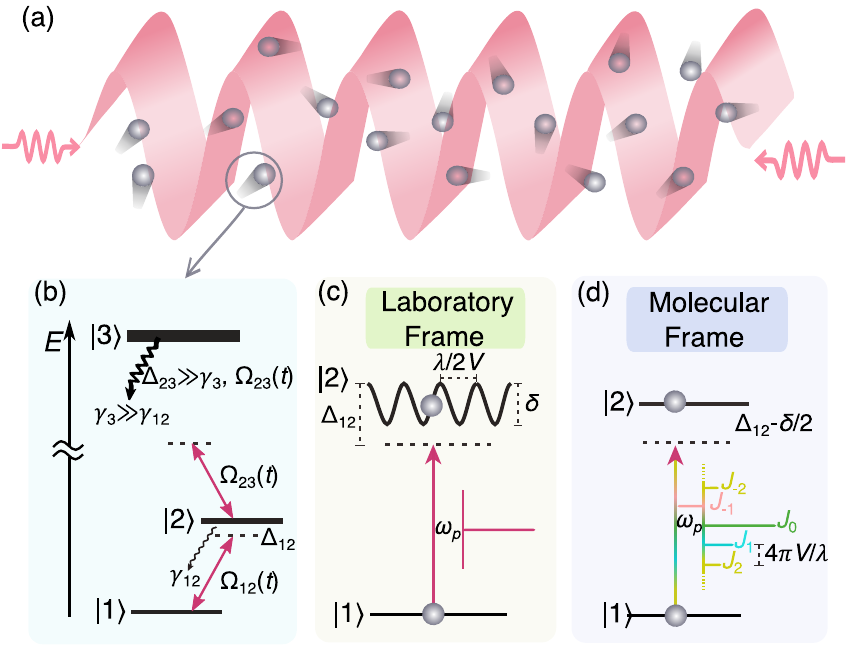}
\caption{Principle of the nonlinear Fano resonance in low-lying states of molecules. 
(a) 
The experimental arrangement for the absorption
spectrum measurement of moving molecules, probed by a standing wave field with wavelength $\lambda$. 
(b) Energy level diagram.
(c)-(d) Simplified energy level diagram for a molecule with speed $V$ in the laboratory frame (c) and in the molecular frame (d). 
}
\label{fig:model}
\end{figure}

\textit{Model and principle.}---The mechanism of NFR is illustrated by 
a conventional precision spectroscopy measurement set-up 
in Fig.~\ref{fig:model}(a),
where a target narrow-linewidth transition ($|1\rangle \leftrightarrow |2\rangle$ with frequency $\omega_{12}$) is probed by a near-resonant laser ($\omega_{p}$). 
However, the probe would not only excite the target transition, 
but also inevitably couple with other excess transitions, 
particularly when a strong probe field is applied in measurements of the ultraweak transition. 
For example, as shown in Fig.~\ref{fig:model}(b), considering an extremely weak (``forbidden'') transition $\left|2\right\rangle \leftarrow \left|1\right\rangle $ and a far-off resonance excited state $\left|3\right\rangle$ which has a very strong transition dipole from $\left|2\right\rangle$, 
the probe laser excites the molecule to both the $|2\rangle $ with a probability $\mathcal{P}\propto\gamma_{12}I/\Delta_{B}^{2}$ and the $\lvert3\rangle$ through the transition $\left|2\right\rangle \leftrightarrow\left|3\right\rangle $ with a Rabi frequency $\Omega_{23}\propto\sqrt{I\gamma_{23}}$. 
Here, $I$ is the laser intensity, $\Delta_{B}$ is the linewidth due to transit-time broadening, which is usually much larger than the natural linewidth of a narrow transition, and $\gamma_{12(23)}$ denotes the spontaneous emission rate from $\left|2(3)\right\rangle $ to $\left|1(2)\right\rangle$~\cite{sm}. \nocite{Gordon2017}
\nocite{Breuer2002}
\nocite{MPI_VUV}
\nocite{Lu2015Apr}
\nocite{Yuan_108nm}
\nocite{Grebenshchikov2013Jun}
\nocite{Abgrall2006_AA}
For the far-off resonance, 
$\left|\Delta_{23}\right|=\left|\omega_{p}-\omega_{23}\right|\gg\Omega_{23},\gamma_{23}$, the excess state induces a modification of the frequency of $\left|2\right\rangle $ due to the AC stark effect~\cite{Foot2004Nov}, which can be estimated as
\begin{equation}
\delta=\frac{\Omega_{23}^{2}}{\Delta_{23}}
 \approx\frac{\gamma_{23}}{\gamma_{12}}
  \frac{\Delta_{B}^2}{\Delta_{23}}
  \mathcal{P}.
\label{eq:delta}
\end{equation}
It is anticipated that for a long-lived energy level $\lvert2\rangle$ with 
$\gamma_{12}\ll\Delta_{B}$, the excess
effect should not be ignored if the amplitude of $\delta$ is comparable with $\Delta_{B}$. 
For example, 
considering a typical requirement of $\mathcal{P}\sim\mathcal{O}\left(10^{-4}\right)$ to enable an observable saturated absorption of a narrow transition,
if $\Delta_{B}/2\pi\sim10^{6}\,\mathrm{Hz}$ for room-temperature gases, an excess transition has $\gamma_{23}/2\pi\sim10^{8}\,\mathrm{Hz}$
and $\Delta_{23}/2\pi\sim10^{15}\,\mathrm{Hz}$ (such as a strong electronic transition approximately 100~nm), 
then the excess effect beyond the conventional two-level
approximation should be considered when $\gamma_{12}/2\pi\lesssim 10^{-4}$~Hz.
The above parameters are consistent with those of the 1.4~$\mu$m narrow transition and the 110~nm electronic dipole transition of HD~\cite{Fantz2006ADNDT-H2, Hua2019}. 

As shown in Fig.~\ref{fig:model}(c), 
when molecules travel through the standing-wave field, periodic excitation and periodic energy level modulation on $\lvert 2\rangle$ appear simultaneously. Thereafter, the system is described by the Floquet Hamiltonian
\begin{align}
H/\hbar= & \sum_{j}\left\{\left[-\Delta_{12}+\delta\cos^{2}\left(k_{p}z_{j}\right)\right]\sigma_{22}^{j} \right.\nonumber \\
 & \left. + \Omega_{12}\cos(k_{p}z_{j})\left(\sigma_{12}^{j}+\sigma_{21}^{j}\right)\right\}.\label{eq:FullModel}
\end{align}
Here, superscript or subscript $j$ indicates the $j$-th molecule, $\Delta_{12}=\omega_{p}-\omega_{12}$, $k_p$ is the wave vector of the probe field, $z_j=V_{j}t$ is the instant location of the molecule with longitudinal velocity $V_j$, $\sigma_{ab}=\left|a\right\rangle \left\langle b \right|$, and $\Omega_{12}$ is 
the Rabi frequency. 
Intuitively, the Floquet interaction 
in the laboratory frame {[}Fig.~\ref{fig:model}(c){]}
can be converted to 
the interaction in the molecular frame {[}Fig.~\ref{fig:model}(d){]} and the corresponding Hamiltonian becomes $
H/\hbar=\sum_{j}\big\{ (-\Delta_{12}+\frac{\delta}{2})\sigma_{22}^{j}+
\Omega_{12}\cos(k_{p}z_{j})\sum_{m}\left[ J_{m}\left(-\zeta_j\right)e^{i2mk_{p}V_{j}t}\sigma_{12}^{j}+h.c\right]\big\}$, where $J_{m}(\cdot)$ is the Bessel function of order $m$, 
$\zeta_j=\delta\lambda/8\pi V_{j}$~\cite{sm}. 
Therefore, moving molecules
in the standing-wave probe field can be treated as static molecules probed by a comb, which is significantly different from the conventional two-level treatment of the system in Fig.~\ref{fig:model}(a).
\begin{figure}
\centering{}\includegraphics[width=\columnwidth]{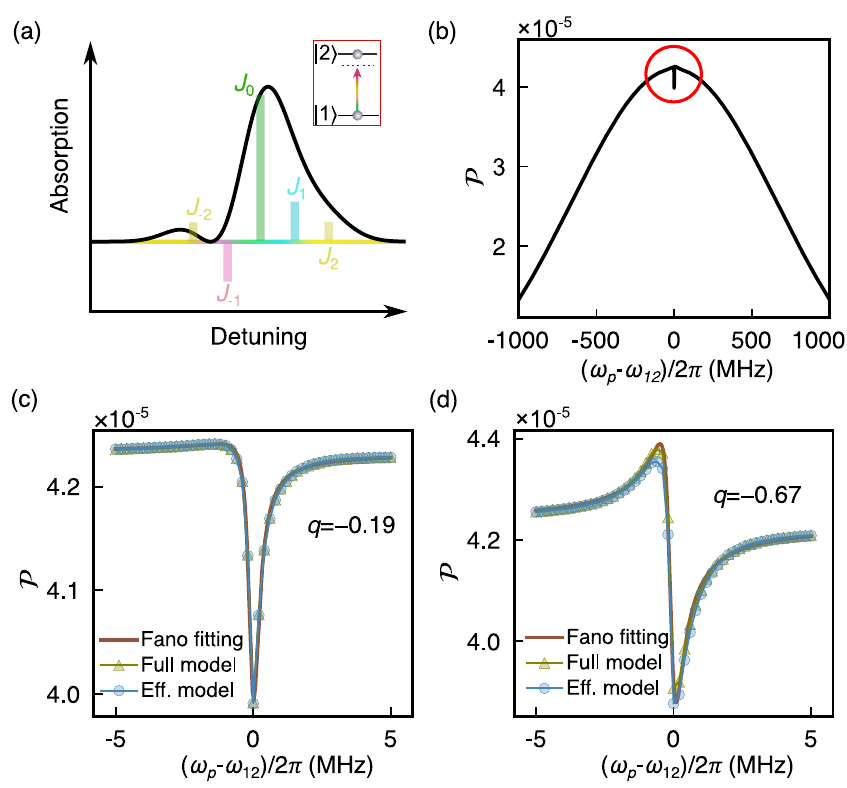} \caption{
(a) Schematic of an asymmetric absorption profile of a static molecule obtained by scanning the probe laser 
in the molecular frame.
(b) Numerically calculated absorption spectrum for a molecule ensemble interacting with a standing-wave field. 
(c)-(d) Detailed absorption features around the line center evaluated by the full model (triangle points) and effective model (circular points), with modulation of $\delta/2\pi=-0.02$ and $-0.12$~MHz, respectively. 
with Fano factors $q$ of $-0.19$ and $-0.67$, respectively. }
\label{fig:SAS}
\end{figure}

\begin{figure}
\centering{}\includegraphics[width=\columnwidth]{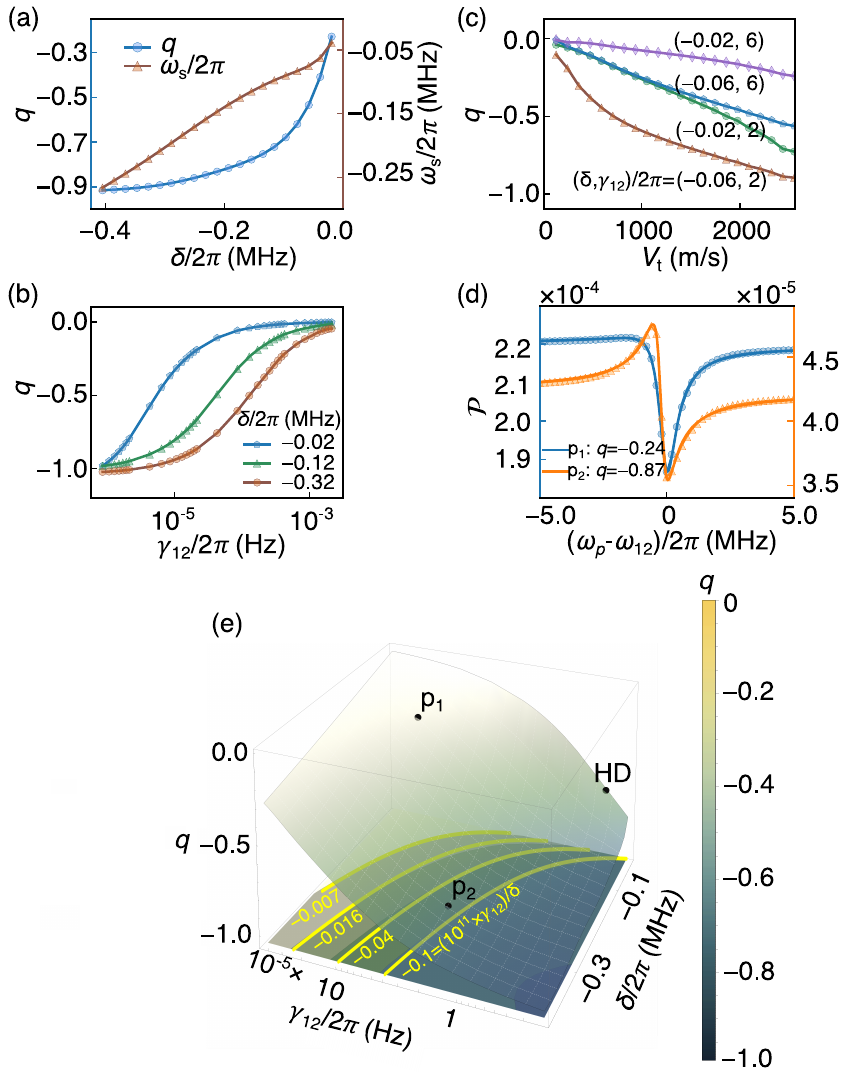} 
\caption{
(a) $q$ and the shift of the fitted center frequency $\omega_{s}/2\pi$ versus the modulation $\delta/2\pi$. 
(b) $q$ versus $\gamma_{12}/2\pi$ for different modulation amplitudes. 
(c) $q$ versus transverse speed $V_{t}$ with a beam waist width of 0.5~mm for the standing-wave field. The label of curves ($\delta/2\pi$~(MHz),$\gamma_{12}/2\pi$~($10^{-5}$~Hz)) indicates the modulation and linewidth of $\lvert2\rangle$.
(d) Absorption spectra of the noted points in (e).
(e) $q$ versus $\delta/2\pi$ and $\gamma_{12}/2\pi$. 
Points p$_{1}$, p$_{2}$ and HD are located at $\delta/2\pi=-0.12, -0.32, -0.022$~MHz and $\gamma_{12}/2\pi=1.3\times10^{-4}, 2.0\times10^{-5}, 3.42\times10^{-6}$~Hz, 
respectively. 
Unless specified, all parameters are the same as those in Fig.~\ref{fig:SAS}.
\label{fig:theory} }
\end{figure}

\textit{Asymmetric spectral profile.}---Although the above discussions hold for all types of absorption spectroscopy,
we focus on atoms or molecules in a gas cell.
For $\lvert\zeta_{j}\rvert\ll1$, only the low order sidebands ($J_{0,\pm1,\pm2}$) of the comb are considered. 
As schematically shown in Fig.~\ref{fig:SAS}(a), an asymmetric absorption spectrum emerges due to the $\pi$ phase difference between sidebands ($J_{\pm1}\left(-\zeta_{j}\right)\approx\mp\zeta_{j}/2$).
Analytically, a Doppler-free fine spectral profile on the Doppler-broadened background,
could be approximately solved as~\cite{sm}
\begin{equation}
    \Delta \mathcal{P}\propto I^2\mathcal{F}^{2}(\Delta_{12})
    +\frac{\sqrt{2}\hbar \omega^3_{12}}{32\pi^{2} c^2 }
    \frac{\delta I\Delta_{B}^{2}}{\gamma_{12}}
    \frac{\partial \mathcal{F}(\Delta_{12})}
    {\partial\Delta_{12}}.
\label{eq:TwoSideband}
\end{equation}
Here, $\mathcal{F}(\Delta_{12})=\mathrm{exp}\left[{-(\Delta_{12}-\frac{\delta}{2})^{2}}/\Delta_{B}^{2}\right]$, and $c$ is the speed of light. 
The first term corresponds to the \textit{symmetric} saturated absorption spectrum (SAS), and the second term indicates an additional \textit{anti-symmetric} spectral profile due to NFR. 
Roughly, the asymmetry parameter $\tilde{q}$ can be evaluated as the ratio of the asymmetric term to the symmetric term~\cite{sm}:
\begin{equation}
    \Tilde{q}\approx  \frac{\sqrt{2}\hbar \omega^3_{12}}{32\pi^{2} c^2 }
    \frac{\delta\Delta_{B}}{\gamma_{12}I}=\frac{3\sqrt{2}\omega_{12}^{3}\gamma_{23}}{4\pi\omega_{23}^{3}\gamma_{12}}\frac{\Delta_{B}}{\Delta_{23}}.
    \label{eq:qtilde}
\end{equation}
Thus, a nonzero $\tilde{q}$ value indicates the appearance of an asymmetric line.
The influence of each parameter on the asymmetry degree is shown explicitly. 
The first term of Eq.~\ref{eq:qtilde} indicates that the asymmetry increases with $\lvert\delta/\gamma_{12}\rvert$ and $\Delta_{B}$. 
Since the amplitude of the modulation relates to the light intensity, $\delta\propto I$, 
we obtain the second equation in Eq.~(\ref{eq:qtilde}).
This implies that the asymmetry does not change much with the light intensity.

\begin{figure}
\centering{}\includegraphics[width=\columnwidth]{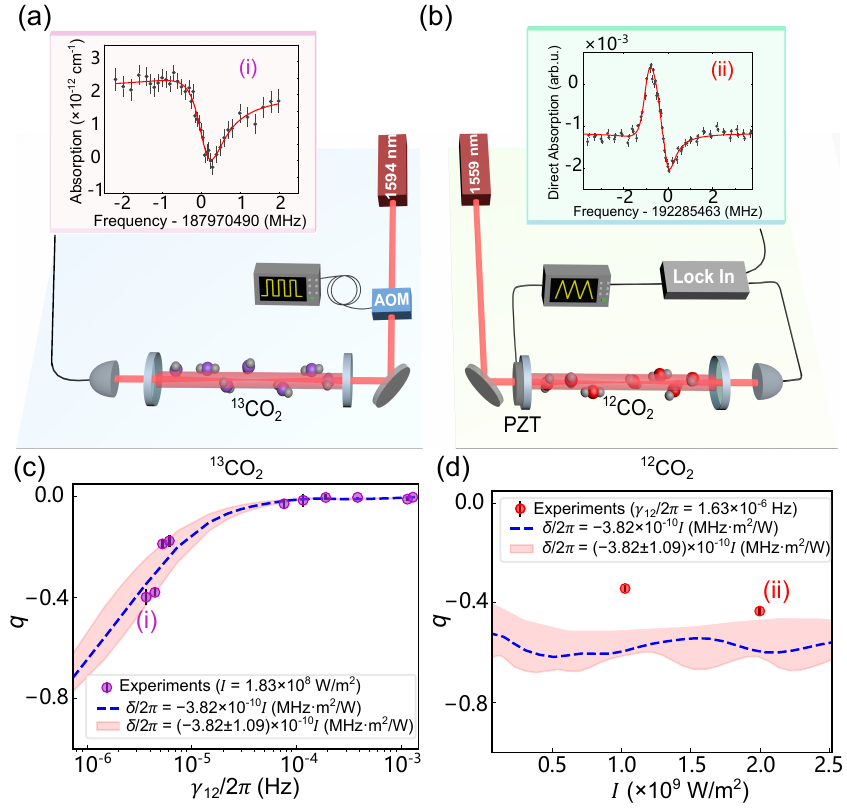}
\caption{
(a) Experimental configuration of cavity ring-down
spectroscopy (CRDS) of $^{13}$C$^{16}$O$_{2}$ with a typical observed spectrum given in the inset.
(b) Experimental configuration of wavelength
modulated cavity enhanced absorption spectroscopy(WM-CEAS) of $^{12}$C$^{16}$O$_{2}$, along with a typical observed spectrum. 
Note that the signal acquired by WM-CEAS essentially presents the first derivative of the normal absorption spectrum and the modulation effect has been included in the fit.
(c) Dependence of $q$ on
$\gamma_{12}/2\pi$ in the $^{13}$CO$_{2}$ measurements (purple circles)
and theory (pink range and blue line) with a standing wave
field intensity of $I=1.83\times 10^8$~W/m$^2$.
(d) Dependence of the $q$ on the standing-wave field intensity $I$ in $^{12}$CO$_{2}$ measurements (red circles) and theory (pink range and blue line) with $\gamma_{12}/2\pi=1.63\times10^{-6}$~Hz.}
\label{fig4}
\end{figure}

When the estimated $\tilde{q}$ value becomes significant ($\gtrsim 0.1$), we should take a numerical calculation with Eq.~(\ref{eq:FullModel}) to investigate the line profile more precisely and further verify the proposed mechanism of NFR.  
In Fig.~\ref{fig:SAS}(b), the typical Doppler-broadened spectrum for molecule ensemble coupling with a standing wave is simulated by taking $\gamma_{12}/2\pi=2\times10^{-5}\,$Hz, $\delta/2\pi=-0.02\,\mathrm{MHz}$, 
and the most probable 
transit-time broadening is $\Delta_B/2\pi=0.4\,$MHz. The Doppler-free profile of Fig.~\ref{fig:SAS}(b) is magnified and shown in Fig.~\ref{fig:SAS}(c), which reveals the asymmetric spectral feature. 
If we change the modulation amplitude of $\delta/2\pi$ from 0.02~MHz to 0.12~MHz, 
the asymmetric feature becomes even more significant [Fig.~\ref{fig:SAS}(d)].
The simulated spectrum could be effectively fit by the Fano function~\cite{Fano_1961_PhysRev}, and we can use the Fano factor $q$ to quantify the degree of asymmetry: 
\begin{equation}
\mathcal{P}=A\frac{q^{2}-1}{1+(\frac{\omega_{p}-\omega_{12}+\omega_{s}}{\Gamma})^{2}}+2A\frac{q(\frac{\omega_{p}-\omega_{12}+\omega_{s}}{\Gamma})}{1+(\frac{\omega_{p}-\omega_{12}+\omega_{s}}{\Gamma})^{2}}+B,
\label{eq:normalFano}
\end{equation}
where $A$, $B$, $\omega_{s}$ and $\Gamma$ are the fitting parameters. 
In particular, $\omega_\mathrm{s}$ corresponds to a fitting frequency shift with respect to the unperturbed transition frequency, and $\Gamma$ represents the linewidth. 
In Figs.~\ref{fig:SAS}(c) and \ref{fig:SAS}(d), the Doppler-free spectra are perfectly fitted by the Fano function, with $q=-0.19$ and $-0.67$, respectively. 
The physics unraveled by the asymptotic solution [Eq.~(\ref{eq:TwoSideband})] is also numerically examined with an effective model that only includes low-order sidebands $J_{0,\pm1}$. 
Although the dynamics of the molecular ensemble are complicated, 
the results of the effective model 
agree well with the full model 
thus further validating the underlying mechanism of NFR. 

The mechanism of NFR and the predictions of $\Tilde{q}$ are further tested by systematically studying the Fano factor via the full model, and the numerical results are summarized in Fig.~\ref{fig:theory}. Apparently, $q$ monotonously increases with $\delta$ but decreases with $\gamma_{12}$, as shown in Fig.~\ref{fig:theory}(a) and (b). Moreover, the fitted frequency shift $\omega_{s}$ in Fig.~\ref{fig:theory}(a)
implies a systematic deviation in the transition frequency
when probing the weak transition. 
Since $\Delta_{B}$ varies with the transverse speed ($V_{t}$) of the molecule, the dependence of $q$ on $V_{t}$ is investigated 
and shown in Fig.~\ref{fig:theory}(c).
These numerical results of $q$ suggest a linear dependence on $\delta$, $\Delta_{B}$, and $1/\gamma_{12}$ when $\lvert q\rvert\ll1$, and agree well with the prediction 
of $\Tilde{q}$. Such a linear dependence of $q$ on $\delta/\gamma_{12}$ is also confirmed in  Fig.~\ref{fig:theory}(e), which shows that the contours of $q$ almost overlap with those of $\delta/\gamma_{12}$. 
Furthermore, typical absorption profiles for the noted points (p$_1$, p$_2$) in Fig.~\ref{fig:theory}(e) are shown in Fig.~\ref{fig:theory}(d), and the spectrum drastically changes from weakly asymmetric to almost anti-symmetric when $q$ increases from $-0.24$ to $-0.87$. 
It is also found that $\lvert q\rvert$ saturates to $1$ for increasing $\lvert\delta\Delta_{B}/\gamma_{12}\rvert$.
The character of $\lvert q\rvert\leq1$ manifests a fundamental difference between the asymmetric profile of our NFR with the conventional Fano resonance~\cite{Fano_1961_PhysRev} because the asymmetry arises from the modulation-induced anti-symmetric components in Eq.~(\ref{eq:TwoSideband}). Therefore, by including the saturation behavior of the fitted $q$ in practice, we can provide a rule of thumb for observing the Fano profiles of weak transitions as $q\approx\Tilde{q}/(1+\lvert\Tilde{q}\rvert)$.
According to this formula, the $q$ values for the spectra in Figs.~\ref{fig:SAS}(c) and \ref{fig:SAS}(d) are -0.16 and -0.54, and those for p$_1$ and p$_2$ in Fig.~\ref{fig:theory}(d) are -0.15 and -0.76, respectively.
We can see that the above formula provides a reasonable estimation of the Fano factor, and more details are given in 
~\cite{sm}.

\textit{Experiment.}---The proposed mechanism could explain the asymmetric profile observed in the cavity-enhanced absorption spectroscopy of HD~\cite{TAO_2018_PRL,Cozijin_2018_PRL, sm}. 
However, the presence of splitting due to the hyperfine structure in the HD molecule~\cite{Komasa2020PRA-HD-Hyperfine}, which is comparable to the observed line width, complicates the interpretation of the asymmetric line profile. 
To circumvent the ambiguity, we further tested the NFR in two independent experiments by measuring the Doppler-free absorption spectra of a series of transitions of CO$_{2}$. The CO$_{2}$ molecule is chosen because its energy structure is well-understood and the hyperfine splitting of the vibration-rotational levels in CO$_{2}$ is also negligible.
Details about the target narrow lines experimentally investigated are given in Supplemental Materials~\cite{sm}.
These lines are 
all well isolated (see Supplemental Figure 1~\cite{sm}),
thus the asymmetric line profile due to interactions from other near-resonance energy levels could be excluded.
By substituting the respective parameters of CO$_2$ into Eq.~(\ref{eq: delta}) or Eq.~(\ref{eq:qtilde}), we predicted that an asymmetric line profile can be observed experimentally when $\gamma_{12} /2\pi \lesssim 10^{-5}$~Hz (see the parameters and prediction in Supplemental Materials~\cite{sm}). 

The first setup, as shown in Fig.~\ref{fig4}(a), employs the cavity ring-down spectroscopy (CRDS) method~\cite{Wang2017}, which measures absorption coefficients for several weak transitions of $^{13}$C$^{16}$O$_{2}$, and an example spectrum is shown in the inset. 
Infrared transitions with $\gamma_{12}$ values spanning the range of $10^{-5} - 10^{-2}$~Hz, weaker than typical CO$_2$ electronic transitions by over one billion times, are investigated under the same experimental conditions. 
All these selected weak transitions are confirmed to be well isolated, therefore possible distortion of the profile due to nearby transitions is excluded.
The extracted $q$ values are shown in Fig.~\ref{fig4}(c). 
Quantitative agreement between the experimental and theoretical results is achieved by considering the AC stark shift  
$\delta/2\pi= (-3.82\pm 1.09)I\times10^{-10}$~MHz$\cdot$m$^2$/W, i.e. 
-0.07$\pm0.02$ ~MHz at a standing wave field intensity of $I=1.83\times10^{8}$~W/m$^{2}$, as noted in the pink area of the figure.

The second setup implements the wavelength-modulated cavity enhanced absorption spectroscopy (WM-CEAS) method~\cite{Hua2019}, which is shown in Fig.~\ref{fig4}(b). 
Transitions of the main isotopologue $^{12}$C$^{16}$O$_{2}$ are studied.
Note that neither the $^{12}$C nor the $^{16}$O nucleus has nuclear spin, and the $^{12}$C$^{16}$O$_{2}$ molecule has no hyperfine structure.
The inset of Fig.~\ref{fig4}(b) shows the WM-CEAS spectrum of the $^{12}$C$^{16}$O$_{2}$ transition at 6413.9526~cm$^{-1}$ with 
an Einstein $A$-coefficient of $1\times10^{-5}$~s$^{-1}$,
corresponding to $\gamma_{12}/2\pi=1.6\,\mathrm{\mu Hz}$. 
Different intracavity optical powers up to 1570~W were used in this experiment, allowing us to reveal the dependence of $q$ on $I$, as shown in Fig.~\ref{fig4}(d).
{We can find that the experimental results agree reasonably with our numerical simulation for $\delta/2\pi=(-3.82\pm 1.09)I\times10^{-10}$~MHz$\cdot$m$^2$/W.
}
The result implies the relation that $\delta$ is proportional to $I$ and $\Tilde{q}$ is insensitive to $I$, which further verifies the prediction of $\Tilde{q}$. 
It is worth noting that the same {relation  between $\delta$ and $I$} was used in simulations of both $^{13}$C$^{16}$O$_{2}$ and $^{12}$C$^{16}$O$_{2}$ spectra, since electronic states are the same for both isotopologues.
The agreement between the theory and experimental results validates our model and confirms the mechanism of NFR. 

\textit{Conclusion.}---We theoretically proposed and experimentally demonstrated a new type of nonlinear Fano-like resonance without involving a continuum. Such a mechanism can significantly change the shape of the spectrum, 
appealing for more careful evaluation and investigation of the potential impact of largely off-resonance strong transitions when performing spectral measurements on ultranarrow resonances. Meanwhile, our work provides an approach to evaluate such a systematic shift for acquiring a more accurate line center {(discussion of the systematic shift for CO$_2$ can be found in 
Supplementary Materials~\cite{sm})}.
Moreover, the observed Fano factor could serve as an indicator of perturbation from unknown strong transitions that could not be effectively probed. 
Note that parameters for distant off-resonance transitions used in our calculation are typical for most electronic dipole transitions of molecules, and the answer to ``when the NFR effect needs to be considered?'' could be referenced to our rule-of-thumb estimate given in Eq.~(\ref{eq:delta}) and Eq.~(\ref{eq:qtilde}).
For the situation that the near-resonance target line is not well isolated, the two-level approximation is not valid and such a system should be solved by a more complex model considering the full multi-level structure.
Ultimately, the results 
may affect a variety of important applications using these narrow transitions, including  optical clocks~\cite{Seiferle_2019_Nature,Brewer_2019_PRL}, long-lived memory of quantum information~\cite{Zhao2009Feb,Lvovsky2009Dec}, testing of quantum theory~\cite{TAO_2018_PRL,Cozijin_2018_PRL}, determination of fundamental constants~\cite{Patra_2020_Science_HD} and searching for new physics beyond the Standard Model~\cite{Safronova_2018_RevModPhys}.

\section*{acknowledgments}
\begin{acknowledgments}
The authors thank F. Meng from NIM and Y.-Y. Zhang from NTSC (CAS) for assistance with the OFC operation.
S.M.H. thanks X.-M. Yang from SUSTech and Y. Luo from USTC for instructive discussions.
This work was
jointly supported by the National Key R\&D Program of China (Grant No.2017YFA0304504),
by the National Natural Science Foundation of China (21688102, 41905018,
21903080, 11922411, U21A6006 and U21A20433), by the Chinese Academy of Sciences (XDB21020100, XDB21010400, XDC07010000), and by the Innovation Program for Quantum Science and Technology (2021ZD0303102). 
C.L.Z. was also supported
by the Fundamental Research Funds for the Central Universities. 
The numerical calculations in this work were performed on the supercomputing system in the Supercomputing Center of the University of Science and Technology of China.
\end{acknowledgments}


%

\end{document}